\documentclass[aps,twocolumn,showpacs,preprintnumbers,amsmath,amssymb,pre,showkeys]{revtex4-1}
\usepackage{bm}
\usepackage{graphicx}

\begin{document}

\title{Entropy and Thermodynamic second laws: New perspective \\
- stochastic thermodynamics and fluctuation theorems}

\author{Hyunggyu Park}
\affiliation{School of Physics and QUC, Korea Institute for Advanced Study, Seoul 02455, Korea}

\date{\today}

\begin{abstract}
Recently, there has been a considerable progress on the issue of the thermodynamic second law, which is known as the law of entropy increase or irreversibility. In particular, a novel symmetry known as the Gallavotti-Cohen symmetry is found in nonequilibrium (NEQ) fluctuations, which leads to so-called fluctuation theorems. The thermodynamic
second law is a simple corollary of fluctuation theorems, from which one can predict quantitatively
how often NEQ processes violate the law of entropy increase. Violations disappear in the thermodynamic limit, but can be observed reasonably well in small systems. In this article, we will briefly introduce the stochastic thermodynamics and derive various fluctuation theorems, including the total entropy production (EP), the work-free-energy relation, the excess and house-keeping EP, and the information entropy.
\end{abstract}



\maketitle

\section{Introduction}
Macroscopic systems tend to reach a stable and quiescent state without any external impulse, which is known as equilibrium (EQ).
Equilibrium has been well explored over more than a century since the key thermodynamic quantity known as {\em entropy}
could be microscopically evaluated by statistical mechanics, thanks mostly to Ludwig Boltzmann. However, most dynamic processes in nature are
non-equilibrium (NEQ) processes and even some steady states (NESS) cannot be described by the EQ Boltzmann distribution. Until recently, not much things are known in general for these NEQ processes and states, except that the EQ Boltzmann entropy in the total system can not decrease between two dynamically connected EQ states (traditional thermodynamic second law) and the fluctuation dissipation theorems from
the linear response theory near EQ.

 The first and most striking breakthrough appeared in 1990's by Evans, Cohen, and Morriss~\cite{evans}, followed by a precise mathematical proof by Gallavotti and Cohen~\cite{gallavotti} on the fluctuation theorem (FT) of the  entropy production in a NEQ steady state. As implied by the name of the FT, it provides an important and universal information on the fluctuation and distribution of the entropy production and, moreover, the thermodynamic second law follows automatically as a corollary of the FT.

 Right after this breakthrough, various forms of the FT's for general stochastic Markovian NEQ processes have been rigorously derived by many researchers, including  Jarzynsky~\cite{jarzynski}, Kurchan~\cite{kurchan}, Crooks~\cite{crooks}, Lebowitz and Spohn\cite{lebowitz}. Sources for NEQ processes
 are also diverse, including time-dependent Hamiltonians and non-conservative driving forces.
 Since then, there have been many interesting generalizations such as FT's for parts of the entropy production~\cite{sasa,speck}, inclusion of information entropy~\cite{sagawa}, odd-parity problem~\cite{spinney,hklee,ckwon1,yeo}, and so forth.  These theoretical results have been confirmed by  experiments on small systems where fluctuations can be observed reasonably well~\cite{bustamante1,bustamante2,ciliberto1,ciliberto2,pak}. We emphasize that all FT's hold regardless of how far the process is from EQ.

 There were also attempts to generalize these ideas to systems with
 a strong coupling to environment~\cite{seifert1}, non-Markovian processes~\cite{speck2}, non-Gaussian/correlated/multiplicative noises~\cite{utsumi}, and open (and also isolated) quantum systems~\cite{quantum}, but
there still remain many unresolved issues in these generalizations.

In this article, we will introduce the {\em stochastic thermodynamics} (ST)~\cite{sekimoto,seifert2,esposito}, which provides a key framework for fluctuation calculations with a new perspective on entropy. It will be seen that many fluctuation theorems can be derived
almost trivially in this ST framework.

\section{Stochastic thermodynamics}

Description of microscopic many-body deterministic systems in terms of a few macroscopic variables is one of the formidable task in physics. However, this is possible in EQ by utilizing the equilibrium statistical mechanics assuming the so-called {\em equally likely} postulate or the entropy maximization. Thus, the EQ statistical mechanics provides a direct link between microscopic deterministic systems and macroscopic thermodynamics. In contrast, there is no general postulate in NEQ, which is consistent with experimental findings. Thus, one has to go through {\em coarse-graining} procedures in order to reduce numerous uninterested degrees of freedom in microscopic systems and describe the system at the mesoscopic level with much less degrees of freedom. With some approximations or truncations, one can usually end up with stochastic dynamic equations such as Langevin equations or Master equations, which embraces time irreversibility in themselves.

\begin{figure}
\centering
\includegraphics[width=0.4\textwidth]{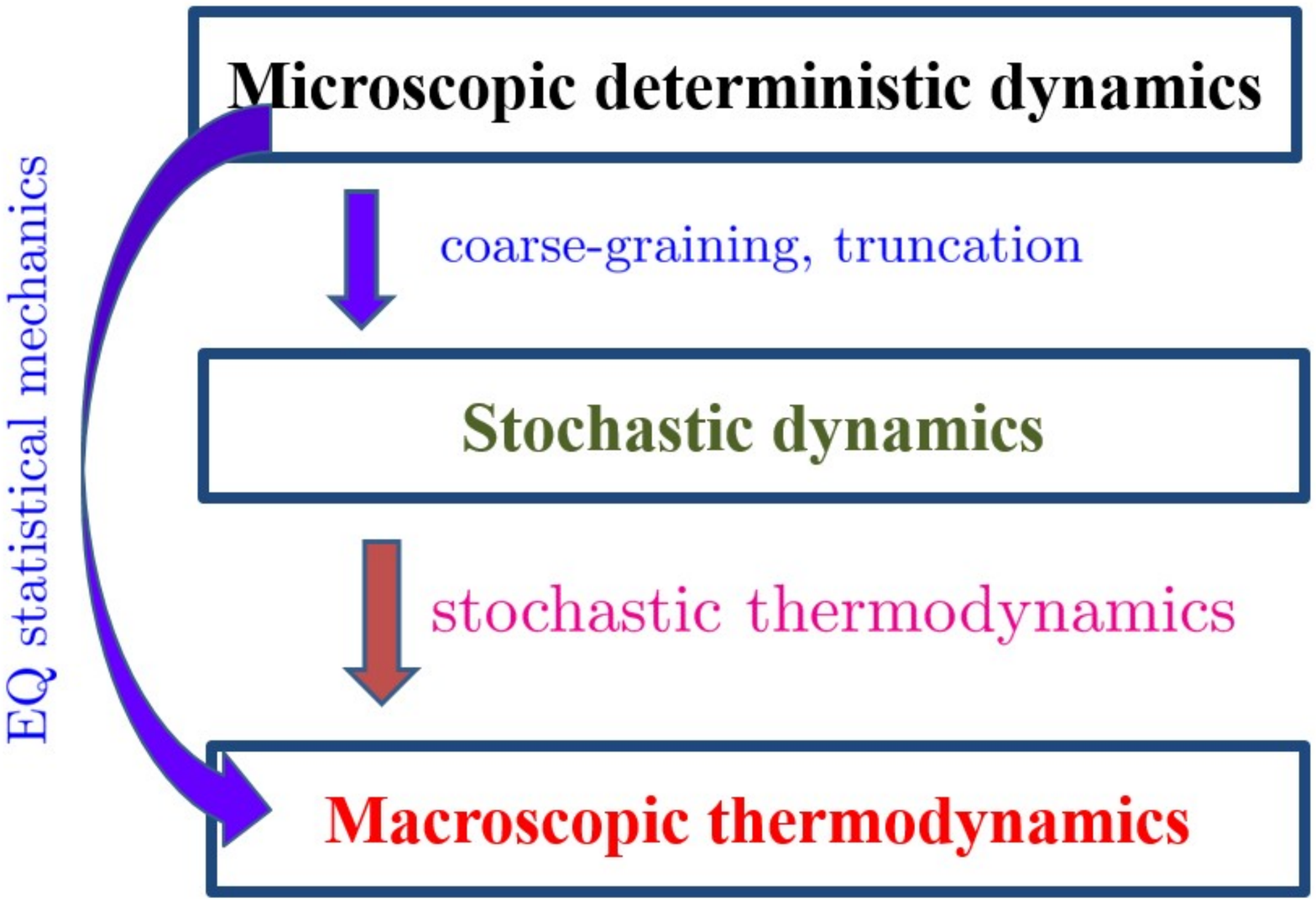}
\caption{(Color online) Conceptual flow diagram to macroscopic thermodynamics in EQ and NEQ. Stochastic thermodynamics bridges from stochastic dynamics described by Langevin or Master equations to macroscopic thermodynamics in general NEQ processes.} \label{fig:flow}
\end{figure}

Stochastic thermodynamics (ST) is a new framework to describe a general NEQ process in terms of the ensemble of dynamic trajectories of a stochastic equation. In particular, thermodynamic quantities such as heat and entropy production along with energy and work are {\em defined} for each dynamic trajectory~\cite{sekimoto,seifert2}. These thermodynamic quantities fluctuate over the ensemble of trajectories and
their ensemble averages become the conventional macroscopic thermodynamic observables. Thus, the ST provides a link between mesoscopic stochastic dynamics to macroscopic thermodynamics. A simple conceptual flow diagram to macroscopic thermodynamics starting from microscopic deterministic dynamics is shown in Fig.~\ref{fig:flow}.

It is useful to look for an analogy between the EQ statistical mechanics and the ST. For convenience, we consider a system
in contact with a single heat reservoir characterized by temperature $T$. EQ concerns only about statics because any macroscopic
thermodynamic observable does not change over time by definition. Nevertheless, system state $q$ fluctuates in the state
space $\{q\}$ with the probability distribution function (PDF) $p(q)$. Note that state $q$ is in principle a very high dimensional object, describing all degrees of freedom of the many-body system. In this article, we only consider
$q=(x,v)$ with position $x$ and velocity $v$ of a single particle in one dimension for simplicity. Generalization to more complex situations is mostly straightforward.

For a given Hamiltonian $H(q)$, it is well known that
the PDF is given as the distribution $p(q)=e^{(F-H(q))/T}$  (canonical ensemble) with the free energy $F=\sum_q e^{-H(q)/T}$ in the Boltzmann unit ($k_B=1$). Then, any state-dependent observable $A(q)$ such as the energy $H(q)$ can be averaged as
$\langle A\rangle=\sum_q p(q) A(q)$. For an isolated system, we have $p(q)=1/\Omega$ (microcanonical ensemble) with the number of {\em accessible} states $\Omega$ for a given macroscopic constraint, which is known as the equally likely postulate.

In the conventional EQ thermodynamics, the system entropy is defined usually by the ensemble-averaged quantity.
One of the important ingredients of the ST is to put the system entropy on the same footing as the energetic quantities~\cite{seifert2}, i.e.~as a fluctuating quantity defined as
\begin{eqnarray}
S(q)\equiv - \ln p(q), \qquad \langle S\rangle =-\sum_q p(q) \ln p(q),\label{entropy}
\end{eqnarray}
where $\langle S\rangle$ is equivalent to the Shannon entropy in the information theory.
With the PDF $p(q)$ inserted for EQ, $\langle S\rangle =\ln \Omega $ for the microcanonical ensemble, which is identical to the Boltzmann entropy, and $\langle S\rangle =-(F-\langle H\rangle )/T$ for the canonical ensemble, consistent with the EQ thermodynamics.
The ST takes the same definition of the system entropy as above even for NEQ.

Now, consider the NEQ process where the dynamics really matters. To describe the dynamics, we consider dynamic trajectories
${\bf{q}}=\{q_t\}$ for $t_1\le t \le t_2$ with the trajectory PDF ${\cal P}({\bf q})$. Any trajectory-dependent observable
$A({\bf q})$ such as the energy change $\Delta H({\bf q})=H(q_{t_2})-H(q_{t_1})$, work $W({\bf q})$ done on the system by an external agent (force), and heat $Q({\bf q})$ dissipated into the reservoir can be averaged over the trajectory ensemble as
$\langle A \rangle =\sum_{\bf q} {\cal P}({\bf q}) A({\bf q})$. Note that fluctuating heat $Q({\bf q})$ in the trajectory space
$\{{\bf q}\}$~\cite{sekimoto} is related to fluctuating work $W({\bf q})$ by the microscopic energy conservation as $W({\bf q})=\Delta H  + Q({\bf q})$.

To discuss the fluctuation theorem and the associated thermodynamic second law, we also need to treat the entropy production (EP) as a fluctuating quantity. In the ST, the EP or equivalently the entropy change of the environment (reservoir) $\Delta S_\text{r} ({\bf q})$ is defined as
\begin{eqnarray}
\Delta S_\text{r} ({\bf q}) = Q({\bf q})/T, \qquad \langle \Delta S_\text{r} \rangle =\langle Q\rangle /T ,\label{EP}
\end{eqnarray}
where $\langle \Delta S_\text{r} \rangle$ is equivalent to the Clausius EP in the conventional {\em reversible} thermodynamics.
As the heat reservoir is assumed to be always in EQ even when the system undergoes a NEQ process,
this result is exactly what is expected. The total entropy change is the sum of the system and reservoir entropy change for a weak coupling to the reservoir as
\begin{eqnarray}
\Delta S_\text{tot} ({\bf q}) = \Delta S({\bf q}) + \Delta S_\text{r} ({\bf q}),\label{tEP}
\end{eqnarray}
where $\Delta S({\bf q})=S(q_{t_2})-S(q_{t_1})=-\ln p_{t_2}(q_{t_2}) + \ln p_{t_1} (q_{t_1})$.
We expect that $\langle \Delta S_\text{tot} \rangle \ge 0$ for any natural process (thermodynamic second law),
but $\Delta S_\text{tot} ({\bf q})$ fluctuates over dynamic trajectories.
So there may be a chance to observe negative total EP in real experiments.
In the next section, we investigate the universal feature of its distribution $P(\Delta S_\text{tot})$ for general
Markovian processes, which is called the fluctuation theorem. The thermodynamic second law is just its corollary.

We end this section with a short summary on the ST. The ST is an ensemble theory for dynamic trajectories with
an arbitrary trajectory PDF, in contrast to the EQ ensemble theory for accessible states with a specific PDF
corresponding to a given environment. In addition, the system entropy and the EP are defined as stochastic
(trajectory-dependent) quantities through Eqs.~\eqref{entropy} and \eqref{EP} along with stochastic heat and work.

\section{Fluctuation theorems}

Fluctuation theorems are based on a very simple probability theory.
We start with this probability theory, introduced by Esposito and Van den Broeck~\cite{esposito}.
Consider two arbitrary normalized PDF's such as
${\cal P}({\bf q})$ and $\tilde{{\cal P}}(\tilde{\bf q})$ with $\sum_{\bf q} {\cal P}({\bf q})=1$ and
$\sum_{\tilde{\bf q}} \tilde{{\cal P}}(\tilde{\bf q})=1$, where variables are linked to each other by
an area-preserving map of $\tilde{\bf q}=\pi ({\bf q})$ (unit Jacobian: $J(\pi)=1$). And we define the so-called {\em relative entropy} as
\begin{eqnarray}
R({\bf q}) \equiv \ln \frac{{\cal P}({\bf q})}{\tilde{{\cal P}}(\tilde{\bf q})},\label{relative}
\end{eqnarray}
which measures the difference between two PDF's. Then, it is trivial to show that
\begin{eqnarray}
\langle e^{-R} \rangle_{\cal P}= \sum_{\bf q} e^{-R({\bf q})} {\cal P}({\bf q})
=\sum_{\tilde{\bf q}} \tilde{\cal P}(\tilde{\bf q})=1,
\end{eqnarray}
where the unit Jacobian is crucial in the derivation. In terms of a variable $R$, this result can be
rewritten as
\begin{equation}
\langle e^{-R} \rangle=\sum_R e^{-R} P(R)=1,
\end{equation}
which constrains the distribution function $P(R)$. This is the {\em integral} FT for the observable $R$.
Utilizing the Jensen's inequality, the above result guarantees
\begin{equation}
\langle R\rangle \ge 0,
\end{equation}
which is called the Kullback-Leibler divergence.

One may get more information on $P(R)$ when a certain condition is satisfied. Consider a functional mapping $f$ such as
$\tilde{{\cal P}}(\tilde{\bf q})=f \circ {\cal P}({\bf q})$, which satisfies the {\em involution} property: $f^2=I$ ($I$: identity mapping).
Then, it is trivial to show that
\begin{eqnarray}
P(R)&=&\sum_{\bf q} \delta(R-R({\bf q})) {\cal P}({\bf q}) \nonumber\\
&=&\sum_{\tilde{\bf q}} \delta(R+{\tilde R}({\tilde{\bf q}})) e^{-{\tilde R}({\tilde {\bf q}})} {\tilde {\cal P}}(\tilde{\bf q})
= {\tilde P}(-R)e^R~ ,
\end{eqnarray}
where we used ${\tilde R}(\tilde{\bf q})= -R({\bf q})$ from the involution property
and ${\tilde P}(R)=\sum_{\tilde{\bf q}} \delta(R-{\tilde R}({\bf q})) {\tilde {\cal P}}(\tilde {\bf q})$.
This relation links two PDF's by reflection with an extra exponential factor $e^R$,
which is called the {\em detailed} FT.

\subsection{Total EP and Irreversibility}

\begin{figure}
\centering
\includegraphics[width=0.4\textwidth]{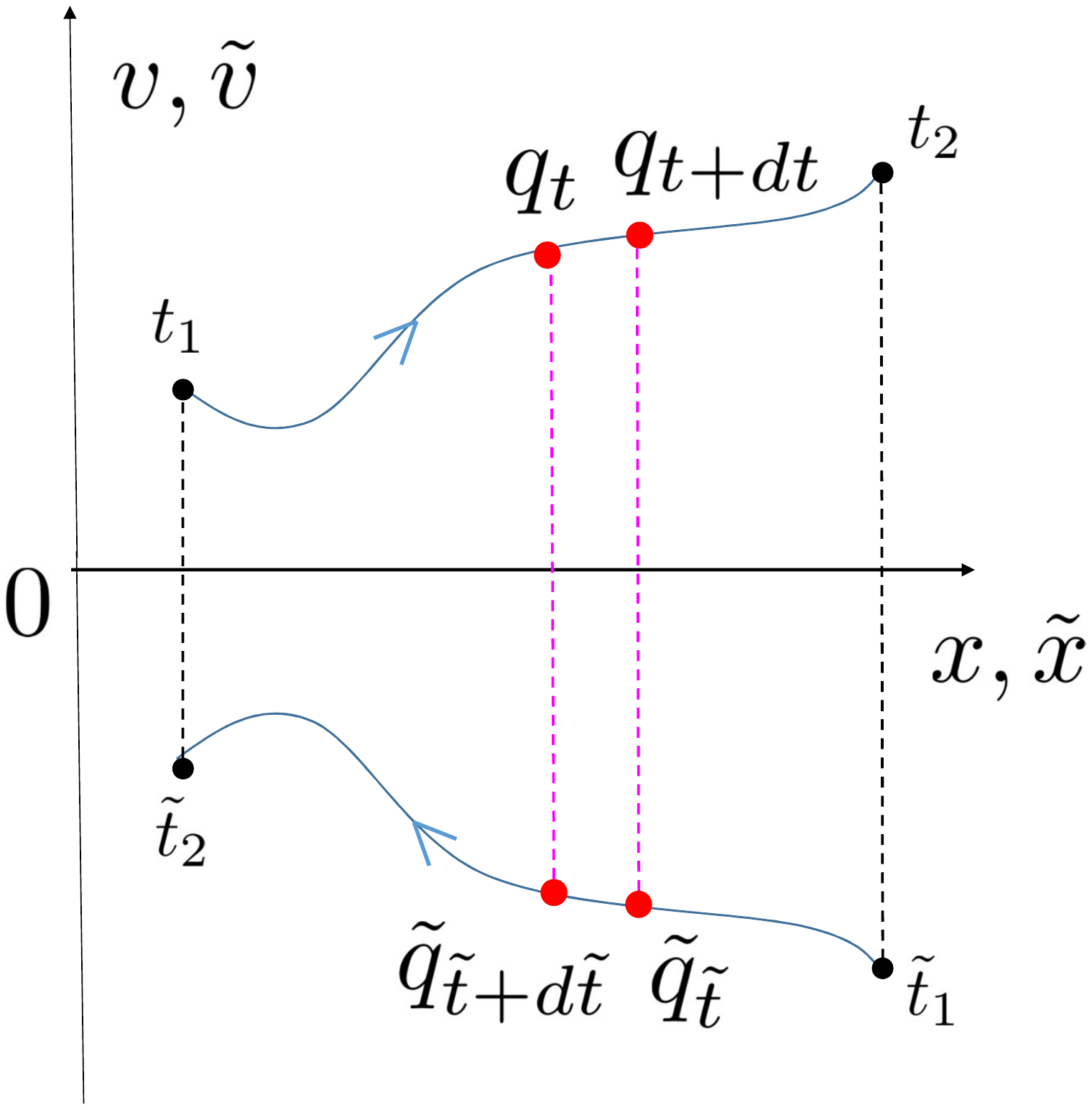}
\caption{(Color online) Dynamic trajectories. Curved lines denote a time-forward
trajectory and its corresponding time-reverse trajectory with ${\tilde q}_{\tilde{t}}=\epsilon q_{t+dt}$ and
${\tilde q}_{\tilde{t}+d\tilde{t}}=\epsilon q_t$ with $\epsilon q=(x, -v)$.
} \label{fig:rev1}
\end{figure}

The total EP measures irreversibility of a given process, so its average vanishes for a reversible process.
In order to define the irreversibility, one should introduce a proper time-reverse process to compare with the original
time-forward process. The standard time-reverse process should obey the {\em same} dynamic equation as in the time-forward process.
A trajectory PDF in the time-reverse process is denoted by ${\tilde {\cal P}}({\tilde{\bf q}})$ for trajectory ${\tilde {\bf q}}
=\{{\tilde q}_{\tilde t}\}$ for ${\tilde t}_1\le {\tilde t} \le {\tilde t}_2$, where all variables in the time-reverse process
are denoted by `${\tilde {\phantom{\tiny{s}}}}$' (Fig.~\ref{fig:rev1}).

It is convenient to consider a trajectory for an infinitesimal time duration between $t$ and $t+dt$. Then, the trajectory PDF
is given as
\begin{eqnarray}
{\cal P}({\bf q})=p_{t}(q_{t}) \Pi_t (q_{t+dt}|q_t), \label{factor}
\end{eqnarray}
where $\Pi_t (q_{t+dt}|q_t)$ is the conditional probability (or propagator)
to reach $q_{t+dt}$ after time $dt$ for a given initial state $q_{t}$.
The propagator function $\Pi_t$  may depend on time $t$ with time-dependent Hamiltonians or driving forces.
The trajectory PDF in the time-reverse process is given as
\begin{eqnarray}
\tilde{\cal P}(\tilde{\bf q})={\tilde p}_{\tilde t}({\tilde q}_{\tilde{t}}) {\tilde \Pi}_{\tilde t} ({\tilde q}_{\tilde{t}+d\tilde{t}}|{\tilde q}_{\tilde{t}}), \label{factor_b}
\end{eqnarray}
for an infinitesimal time duration between ${\tilde t}$ and ${\tilde t}+d{\tilde t}$
with $\tilde{t}=t+dt$ and $\tilde{t}+d\tilde{t}=t$ ($d\tilde{t}=dt$). During this time interval,
both dynamic equations should be identical, so ${\tilde \Pi}_{\tilde t}=\Pi_t$. This implies that the time variation
of the Hamiltonian or driving force for a finite time interval is reversed in the time-reverse process.
For example, if one increases the external force in the time-forward process, then the external force should decrease
in the time-reverse process, which seems natural.

To define the irreversibility for each trajectory, we choose a time-reverse trajectory as the exact reversal of a time-forward trajectory by setting  ${\tilde q}_{\tilde{t}}=\epsilon q_{t+dt}$,
${\tilde q}_{\tilde{t}+d\tilde{t}}=\epsilon q_t$, where $\epsilon$ is the parity operator as $\epsilon q=(x, -v)$,
see Fig.~\ref{fig:rev1}.
Furthermore, we require that the initial probability  for the time-reverse process
is exactly the same as the final probability  for the time-forward process:
${\tilde p}_{\tilde t}({\tilde q}_{\tilde{t}})= p_{t+dt}(q_{t+dt})$. This implies that the time-reverse
process starts right after the end of the time-forward process with the same ensemble except for inverting the velocity direction (note that ${\tilde v}_{\tilde t}=-v_{t+dt}$).

We define the irreversibility of an infinitesimal dynamic trajectory ${{\bf q}}$ as
\begin{eqnarray}
d\text{Irr}({\bf q})\equiv \ln {\frac{{\cal P}({\bf q})}{{\tilde{\cal P}}({\tilde{{\bf q}})}}}, \label{Irr0}
\end{eqnarray}
where the time-reverse trajectory ${\tilde {\bf q}}$ and its initial ensemble are defined as above.
It is trivial to show that the Jacobian of ${\tilde {\bf q}}=\pi ({\bf q})$ is unity. Thus, the integral
FT is automatic by the probability theory as
\begin{eqnarray}
\langle e^{-d\text{Irr}} \rangle =1 \quad \text{and}\quad \langle d\text{Irr} \rangle \ge 0~. \label{Irr}
\end{eqnarray}

It is clear that the involution property holds for the conditional probability;
$f^2\circ \Pi_t (q_{t+dt}|q_t)=\Pi_t (q_{t+dt}|q_t)$. However, it
holds for the initial PDF only in the steady state, guaranteeing ${\tilde p}_{\tilde{t}+d{\tilde t}}
({\tilde q}_{{\tilde t}+d{\tilde t}})=p_t(q_t)$.
As the time-dependent dynamics does  not allow a steady state, the involutarity
is valid in the steady state with a time-independent propagator $\Pi$.
In this case, the detailed FT holds as
\begin{eqnarray}
P(d\text{Irr})=P(-d\text{Irr})e^{d\text{Itt}}~,
\end{eqnarray}
where we drop `${\tilde {\phantom{\tiny{s}}}}$' in the PDF of the right hand side because the time-reverse process
is identical to the time-forward process with the same initial ensemble (steady state) and the same propagator. This reflection symmetry with
a specific exponential weight is generally called as the {\em Gallavotti-Cohen} (GC) symmetry of the PDF.

How is the irreversibility $d\text{Irr}$ related to the thermodynamic quantity such as the total entropy production?
From Eqs.~\eqref{factor} and \eqref{factor_b}, one can see that $d\text{Irr}$ can be divided into two parts: The first
part is the change of the system entropy $dS({\bf q})=-\ln p_{t+dt}(q_{t+dt})+\ln p_t(q_t)$ defined in Eq.~\eqref{entropy} and
the second part is the logarithm of the conditional probability ratio. Here comes the key nontrivial derivation originally
by Schnakenberg~\cite{schnakenberg} and later by many others in general stochastic systems~\cite{kurchan,lebowitz,hinrichsen} as
\begin{eqnarray}
\ln {\frac{{\Pi_t}(q_{t+dt}|q_t)}{{\Pi_t}({\tilde q}_{\tilde{t}+d\tilde{t}}|{\tilde q}_{\tilde{t}})}}
=\frac{dQ({\bf q})}{T}=dS_\text{r} ({\bf q}),\label{Schfor}
\end{eqnarray}
where $dQ({\bf q})$ is stochastic  heat dissipated into the reservoir with temperature $T$.
Thus, the irreversibility is directly related to the EP as
\begin{eqnarray}
d\text{Irr}({\bf q})=dS({\bf q})+dS_\text{r}({\bf q})=dS_\text{tot}({\bf q})~.
\end{eqnarray}
Therefore we finally get the integral and detailed FT's for the total EP as
\begin{eqnarray}
\langle e^{-dS_\text{total}} \rangle=1 \quad \text{and} \quad \frac{P(dS_\text{total})}{P(-dS_\text{total})}=e^{dS_\text{total}}~.
\end{eqnarray}
It is trivial to extend the FT's to any finite time interval, due to the Makovianity of the process.
We emphasize that the integral FT holds for any initial condition, as is so the thermodynamic second law $\langle dS_\text{total}\rangle \ge 0$, but the detailed FT (GC symmetry) only holds in the steady state.

One may guess that this symmetry could be recovered in the infinitely long time limit with any non-steady-state initial condition, because the system reaches its steady state and the initial condition information should be lost in this limit. However, it turns out that this guess is incorrect and the initial memory persists forever in the far tail of the PDF, where the extreme rare events
dominate~\cite{farago,zon,noh1,jslee,kwangmoo}. One may also ask a question whether the integral FT for heat may hold in the steady state as $\langle e^{-dQ/T}\rangle_\text{ss}=1$(?), because the average of the system EP vanishes ($\langle dS\rangle_\text{ss}=0$) and also ask a question whether accumulated heat for a long time in the steady state satisfies the integral FT because accumulated heat should be extremely larger than the system entropy change. Both are not true, simply because $dS$ is a fluctuating quantity which can be very large (unbounded) even with a very small probability. As the integral FT is simply the integral (or sum) of $e^{-dS_\text{total}}$ with the weight of the PDF,  extremely large negative $dS_\text{total}$ contributes to the integral significantly even though its probability
is exponentially small. It is shown that the integral FT for heat can be
satisfied only with the special initial condition of the uniform PDF~\cite{jslee,kwangmoo}.

We will not derive the Schnakenberg formula, Eq.~\eqref{Schfor}, in this article, but only mention that
this formula was rigorously derived for various general stochastic systems, e.g., described by the Langevin equation and the Master equation. One can still make a rough argument inferring from the Arrhenius reaction (hopping) rate such as $e^{-dH/T}$ that the logarthimic ratio of two propagators is
just a energy difference between two states $q_t$ and $q_{t+dt}$ divided by $T$, which should
dissipate as heat into the reservoir.

It is noteworthy to point out that
there are some examples where the Schnakenberg formula breaks down and generates an extra term  besides $dS_\text{r}$.
One well-studied case is when the system is influenced by velocity-dependent forces~\cite{spinney,hklee,ckwon1,yeo,hklee2}.
The Lorentz magnetic force is fundamental and velocity-dependent, but fortunately one can show  the FT's for the total EP
by taking the time-reverse process with the opposite direction of the magnetic field $B$ in Eq.~\eqref{factor_b} as ${\tilde \Pi}_{\tilde t}(B)=\Pi_t(-B)$. For general velocity-dependent
forces, this remedy does not work and one can not avoid an extra EP term in the thermodynamic second law and
the FT's~\cite{ckwon1,GC,noh2}. These phenomenological velocity-dependent forces could be found in many realistic processes including active particle dynamics, cold damping models, information-assisted engines, and so on, where the standard thermodynamic second law is significantly modified~\cite{hklee2,mm,ito,active_Brown,active,granular,cold_damping}.

\subsection{Work and Free energy}\label{wf}

It is usually difficult to measure the EP and heat experimentally. However, work done by the external force can be estimated rather
easily. One can derive the so-called work-free-energy relation almost trivially, by only changing the initial ensemble in the previous discussion. This relation may be very useful and easily checked by experiments and simulations.

We take the same time-forward and time-reverse trajectories as before in Fig.~\ref{fig:rev1} along with Eqs.~\eqref{factor} and \eqref{factor_b}, but with a special choice of the initial conditions as EQ Boltzmann distributions:
\begin{eqnarray}
&&p_t(q_t)=e^{\left(F_t-H_t(q_t)\right)/T} ~, \nonumber\\
&&{\tilde p}_{\tilde t}({\tilde q}_{\tilde{t}})=e^{\left(F_{t+dt}-H_{t+dt}(q_{t+dt})\right)/T}~,\label{work_init}
\end{eqnarray}
where the Hamiltonian $H_t$ may vary with time as well as the free energy defined by its Hamiltonian for a given time.
Note that the final PDF in the time-forward process $p_{t+dt}(q_{t+dt})$ can be arbitrary, depending on the dynamics and the time duration,  which has nothing to do with the initial EQ PDF in the time-reverse process ${\tilde p}_{\tilde t}({\tilde q}_{\tilde{t}})$. In this sense, the trajectory PDF ${\cal {\tilde P}}({\tilde {\bf q}})$ is not
the usual PDF of the time-reverse process.

However, the trajectory probability ratio leads to an interesting relation as
\begin{equation}
\ln {\frac{{\cal P}({\bf q})}{{\tilde{\cal P}}({\tilde{{\bf q}})}}}= \frac{-dF+dH({\bf q})+dQ({\bf q})}{T}=
\frac{dW({\bf q})-dF}{T}, \label{work_ratio}
\end{equation}
where $dF=F_{t+dt}-F_t$, $dH({\bf q})=H_{t+dt}(q_{t+dt})-H_t(q_t)$, and stochastic work $dW({\bf q})=dH({\bf q})+dQ({\bf q})$.
The previous simple probability theory reads
\begin{equation}
\langle e^{-dW/T} \rangle=e^{-dF/T} \quad \text{and} \quad \frac{P(dW)}{{\tilde P}(-dW)}=e^{(dW-dF)/T}~, \label{workFT}
\end{equation}
which are known as the {\em Jarzynsky identity}~\cite{jarzynski} and the {\em Crooks relation}~\cite{crooks}, respectively.
The corresponding thermodynamic second law is $\langle dW\rangle \ge dF$.

Note that the involution property is always satisfied in this case because each initial PDF is given
by the Boltzmann distribution with the Hamiltonian at its initial time. However, the time-reverse
process is not identical to the time-forward process, due to different initial conditions
$p_t \neq {\tilde p}_{\tilde t}$ as in Eq.~\eqref{work_init}.
Furthermore, the time-reverse process for a finite time interval is different from the time-forward process, because
the direction of the time variation in the external force or Hamiltonian is opposite.
Therefore, we can not drop `${\tilde {\phantom{\tiny{s}}}}$'
in the detailed FT in Eq.~\eqref{workFT}.
We also note that  work can be done by a non-conservative time-independent force.
In this case, the Hamiltonian is independent of time, so $p_t= {\tilde p}_{\tilde t}$ and then we can drop `${\tilde {\phantom{\tiny{s}}}}$' along with $dF=0$ in Eq.~\eqref{workFT}.

Finally, we emphasize that
the work free energy FT's in Eq.~\eqref{workFT} holds only when we start with EQ distributions.
So, if one measures the cumulated work starting from EQ, the FT's are valid.
However, both integral and detailed FT's break down for general NEQ processes starting an arbitrary initial PDF, so the average work can become smaller than the free energy difference.

\subsection{Excess and House-keeping EP}

As noticed from the simple probability theory, any fluctuating quantity written as a form of relative entropy in Eq.~\eqref{relative} satisfies the integral FT with unit Jacobian $J$. Even for $J\neq 1$,
a modified integral FT holds as $\langle e^{-R^\prime}\rangle =1$ with $R^\prime = R - \ln J$.
Moreover, one can choose an arbitrary process represented by the trajectory PDF ${\tilde {\cal P}}(\tilde {\bf q})$ (not time-reverse process), where the conditional probabilities $\Pi$  and ${\tilde \Pi}$ in Eqs.~\eqref{factor} and \eqref{factor_b} are not related to each other. Thus, in principle, there are arbitrarily many quantities with valid FT's and so their averages never decrease
in time~\cite{kurchan2}.

What is important is whether these quantities could be identified by physical thermodynamic variables, which can be measured in average by experiments. Here, we mention a couple of examples of such cases. The first example is the so-called {\em excess} EP~\cite{sasa,oono},
$dS_\text{ex}$, which represents entropy (heat) production responsible for
transitions between different steady states plus the system EP. This can be obtained by
taking~\cite{esposito}
\begin{equation}
{\tilde \Pi}_{\tilde t}^* ({\tilde q}_{\tilde{t}+d\tilde{t}}|{\tilde q}_{\tilde{t}})
={\Pi}_{t} ({\tilde q}_{\tilde{t}}|{\tilde q}_{\tilde{t}+d\tilde{t}})
\frac{p_\text{ss}^{(t)} ({\tilde q}_{\tilde{t}+d\tilde{t}})}
{p_\text{ss}^{(t)} ({\tilde q}_{\tilde{t}})}~,\label{excess}
\end{equation}
where $p_\text{ss}^{(t)}$ is the {\em instantaneous} steady-state PDF defined as the expected steady-state PDF
if the time-dependent Hamiltonian is kept unchanged at a given time $t$ such that
$\sum_q \Pi_t(q^\prime|q) ~p_\text{ss}^{(t)}(q)=p_\text{ss}^{(t)}(q^\prime)$.
The stochasticity of this $*$-dynamics is preserved by
$\sum_{{\tilde q}_{\tilde{t}+d\tilde{t}}}{\tilde \Pi}_{\tilde t}^* ({\tilde q}_{\tilde{t}+d\tilde{t}}|{\tilde q}_{\tilde{t}})=1$
from Eq.~\eqref{excess}, which makes ${\tilde{\cal P}}^* ({\tilde{\bf q}})={\tilde p}_{\tilde t}({\tilde q}_{\tilde{t}}) {\tilde \Pi}_{\tilde t}^* ({\tilde q}_{\tilde{t}+d\tilde{t}}|{\tilde q}_{\tilde{t}})$ normalized.

We set ${\tilde q}_{\tilde{t}}=q_{t+dt}$ and ${\tilde q}_{\tilde{t}+d\tilde{t}}=q_t$ with
$\tilde{t}=t+dt$, $\tilde{t}+d\tilde{t}=t$. Notice that the
variable changes between $q$ and ${\tilde q}$ are different from the standard time-reverse transformation
(no parity operator).
With the choice of ${\tilde p}_{\tilde t}({\tilde q}_{\tilde{t}}) =p_{t+dt}(q_{t+dt})$,
we define
\begin{equation}
dS_\text{ex}({\bf q}) \equiv
\ln {\frac{{\cal P}({\bf q})}{{\tilde{\cal P}}^*({\tilde{{\bf q}})}}}=
dS({\bf q})+ \frac{dQ_\text{ex}({\bf q})}{T}, \label{excess_ratio}
\end{equation}
with the  excess heat $dQ_\text{ex}({\bf q})$ as
\begin{equation}
dQ_\text{ex}({\bf q})=T\ln {\frac{p_\text{ss}^{(t)} (q_{t+dt})}{p_\text{ss}^{(t)} (q_{t})}}~.\label{excess_heat}
\end{equation}
With the time-independent propagator $\Pi$, the instantaneous steady state $p_\text{ss}^{(t)}$ does not depend on time $t$ any more.
If one starts with the steady state, then $dS_\text{ex}({\bf q})$ is identically zero for any trajectory ${\bf q}$.
Thus, the excess EP is responsible for evolution of the instantaneous steady states.

With the unit Jacobian again, we get the integral FT automatically as
\begin{eqnarray}
\langle e^{-dS_\text{ex}} \rangle=1 \quad \text{and} \quad\langle  dS_\text{ex}\rangle \ge 1~.
\end{eqnarray}
In the adiabatically slow limit, the system is almost always in the steady state, then $dS_\text{ex}({\bf q})=0$ again for
any ${\bf q}$.
This is why $dS_\text{ex}$ is also called the {\em non-adiabatic} EP~\cite{esposito}. It is interesting to note that
the involution property holds for the conditional probability, but does not hold for the
initial ensemble in general except for the adiabatic limit,where the detailed FT is meaningless due to
$P(dS_\text{ex})=\delta (dS_\text{ex})$.

Another interesting quantity is the so-called {\em house-keeping} EP~\cite{sasa,speck,oono}, $dS_\text{hk}$, which usually represents
entropy (heat) generated in order to maintain the NEQ steady state. This is simply defined as the remaining EP excluding $dS_\text{ex}$
such as
\begin{eqnarray}
dS_\text{hk}({\bf q}) &\equiv& dS_\text{total}({\bf q})-dS_\text{ex}({\bf q}) \label{hk_ratio}\\
&=&\ln {\frac{{\Pi_t}(q_{t+dt}|q_t)p_\text{ss}^{(t)} (q_{t})}{{\Pi_t}(\epsilon q_t|\epsilon q_{t+dt})p_\text{ss}^{(t)} (q_{t+dt})}}=\frac{dQ_\text{hk}({\bf q})}{T}~,\nonumber
\end{eqnarray}
where the house-keeping heat $dQ_\text{hk}({\bf q})=dQ({\bf q})-dQ_\text{ex}({\bf q})$.
Note that $dS_\text{hk} ({\bf q})$ is independent of initial conditions  and is also called the {\em adiabatic} EP~\cite{esposito},
in contrast to $dS_\text{ex}$. We emphasize that
$dS_\text{hk}({\bf q})=0$ for any trajectory ${\bf q}$ for systems without any non-conservative or time-dependent
(NEQ) force, because their steady states should be EQ states and their dynamics satisfy
two (independent) EQ properties such as the detailed balance; $\Pi (q^\prime|q)p_\text{ss}(q)=\Pi (\epsilon q|\epsilon q^\prime)p_\text{ss}(\epsilon q^\prime)$
and the parity symmetry of the PDF; $p_\text{ss} (q)=p_\text{ss} (\epsilon q)$~\cite{parity_eq}.

One may ask whether $dS_\text{hk}$ also satisfies the FT's. This turns out to be correct only in the overdamped dynamics
~\cite{speck}, where all state variables have the even parity ($\epsilon=1$). In this case, we find that the term in Eq.~\eqref{hk_ratio} is
identical to the propagator of the time-forward stochastic $*$-dynamics as
\begin{equation}
{\Pi_t}(q_t| q_{t+dt})p_\text{ss}^{(t)} (q_{t+dt})/p_\text{ss}^{(t)} (q_{t}) ={\Pi}_{ t}^* (q_{t+dt}|q_t)~,
\end{equation}
which leads to $dS_\text{hk} ({\bf q})=\ln [{\cal P}({\bf q})/{\cal P}^*({\bf q})]$
with normalized ${{\cal P}}^* ({{\bf q}})={ p}_{t}({ q}_{{t}}) {\Pi}_{t}^* ({q}_{{t}+d{t}}|{ q}_{{t}})$.
Thus, we get the FTs for the systems only with the even-parity state variables as
\begin{equation}
\langle e^{-dS_\text{hk}} \rangle=1 \quad \text{and} \quad \frac{P(dS_\text{hk})}{{P}^*(-dS_\text{hk})}=e^{dS_\text{hk}}~, \label{hkFT}
\end{equation}
where the initial conditions are arbitrary for both integral and detailed FT's.
With the odd-parity variables like velocities, both FT's for $dS_\text{hk}$ break
down~\cite{spinney,hklee}, but one can divide $dS_\text{hk}$ into two parts which are responsible for the breakage of the detailed balance and for
the parity asymmetry of the steady-state PDF. The FT's hold for the former, but not for the latter~\cite{hklee,yeo}.

For the NEQ dynamics driven by a time-independent non-conservative force
(time-independent propagator $\Pi$), $dS_\text{ex}({\bf q})=0$ for any ${\bf q}$ starting from the steady state
(note that $dQ_\text{ex}({\bf q}) \neq 0$, even though $\langle dQ_\text{ex}\rangle_\text{ss} = 0$).
Therefore, $dS_\text{hk}({\bf q})$ is exactly equivalent to $dS_\text{total}({\bf q})$,
which should satisfy the FT's even with the odd-parity variables. The house-keeping heat
can be written as $dQ_\text{hk}({\bf q})=dQ ({\bf q}) + T dS ({\bf q})$, implying that
the stochastic house-keeping heat differs from the total stochastic heat even in the steady state, but
their steady-state averages are the same: $\langle dQ_\text{hk}\rangle_\text{ss}=\langle dQ \rangle_\text{ss}$.
The thermodynamic second law  also guarantees $\langle dQ_\text{hk}\rangle_\text{ss} \ge 0$.

\subsection{Information entropy }

Finally, we briefly mention how to incorporate information into the FT~\cite{sagawa}. This is almost trivial because
the information is usually expressed by the Shannon entropy as in the stochastic thermodynamics.
The simplest example involving information is a composite system (system + memory), where only the system
is in contact with a thermal reservoir.
The stochastic entropy of the composite system can be decomposed as
\begin{equation}
S_\text{comp}(q,m)=S_\text{sys}({q})+S_\text{mem}({m})-I({q},{ m})~,\label{muinfo}
\end{equation}
where ${ q}$ and ${m}$ denote state variables for the system and the memory, respectively, and
$I$ is the mutual information, representing the correlation between the system and the memory.
Of course, $dS_\text{total}=dS_\text{comp}+dQ_\text{sys}/T$ satisfies the integral fluctuation theorem for any
initial condition.

It is useful to take a special example such as the post-measurement process~\cite{sagawa,ckwon3,um} in the so-called Maxwell's demon problem,
where the memory state does not change in time during the feedback/relaxation process after the measurement.
Furthermore, we start from the EQ initial distribution for the system such as
$p_t(q_t,m_t)=p_t(q_t)p_t(m_t|q_t)$ with $p_t(q_t)=e^{(F_t-H_t(q_t))/T}$
with the system Hamiltonian $H_t$ and
an arbitrary conditional PDF of the memory $p_t (m_t|q_t)$, which is related to the mutual information
as $p_t (m_t|q_t)=p_t(m_t) e^{I_t(q_t,m_t)}$ by Eq.~\eqref{muinfo}.

Now, we take the similar methodology as in~\ref{wf}
with ${\tilde p}_{\tilde t}({\tilde q}_{\tilde{t}},{\tilde m}_{\tilde{t}})=
{\tilde p}_{\tilde t}({\tilde q}_{\tilde{t}}) {\tilde p}_{\tilde t}
({\tilde m}_{\tilde{t}}|{\tilde q}_{\tilde{t}})$ by setting
${\tilde q}_{\tilde{t}}=\epsilon q_{t+dt}$, ${\tilde q}_{\tilde{t}+d\tilde{t}}=\epsilon q_t$,
${\tilde m}_{\tilde{t}}= \epsilon m_{t}$, ${\tilde m}_{\tilde{t}+d\tilde{t}}= \epsilon m_t$,
with $\tilde{t}=t+dt$, $\tilde{t}+d\tilde{t}=t$.
We choose
${\tilde p}_{\tilde t}({\tilde q}_{\tilde{t}}) =e^{\left(F_{t+dt}-H_{t+dt}(q_{t+dt})\right)/T}$
and ${\tilde p}_{\tilde t}
({\tilde m}_{\tilde{t}}|{\tilde q}_{\tilde{t}})={\tilde p}_{\tilde t}
({\tilde m}_{\tilde{t}}) e^{{\tilde I}_{\tilde t}({\tilde q}_{\tilde t},{\tilde m}_{\tilde t})}$
with ${\tilde p}_{\tilde t} ({\tilde m}_{\tilde{t}}) =p_t(m_t)$
and ${\tilde I}_{\tilde t}({\tilde q}_{\tilde t},{\tilde m}_{\tilde t})=I_{t+dt} (q_{t+dt}, m_t)$.
Then, we get the integral FT only when the system starts from EQ as
\begin{equation}
\langle e^{-dW/T+dI}\rangle=e^{-dF/T}~,\label{work_info}
\end{equation}
where $dW=dH+dQ_\text{sys}$ and $dI=I_{t+dt} (q_{t+dt},m_t)-I_t(q_t,m_t)$.
The corresponding second law is $\langle dW\rangle \ge T\langle dI\rangle +dF$.
Without getting help from the free energy difference (in the case of $dF=0$), one can
utilize the information gained in the measurement process for the work extraction
(negative $\langle dW\rangle$) during the feedback/relaxation process where
the mutual information (correlation) decreases as $\langle dI\rangle <0$.
This is the key idea of an information engine.

\section{Concluding remarks}

In this article, we focus on the FT's and thermodynamic second laws for classical Markovian dynamics with a single heat reservoir.
There have been some progresses in various generalizations such as
multiple reservoirs~\cite{esposito2,jslee2}, quantum systems~\cite{quantum}, strong coupling to environments~\cite{seifert1}, non-Markovian processes~\cite{speck2}, and
non-thermal noises~\cite{utsumi}. We believe that the thermodynamic second law for the total EP would be valid
for any fundamental and natural phenomenological processes. Thanks to the stochastic thermodynamics,
we can prove the fluctuation theorems easily for some systems, which shed a novel light on the directional mystery of
the macroscopic time arrow. However, there are still many realistic systems waiting
for being explored.

\begin{acknowledgments}
We thank many colleagues who contributed to some works presented here and also for useful discussions. In particular,
Chulan Kwon, Jae Dong Noh, Joonhyun Yeo, Hyun Keun Lee, Jaegon Um, Jae Sung Lee, and Haye Hinrichsen are deeply appreciated.
This research was supported by the NRF Grant No.~2017R1D1A1B06035497.

\end{acknowledgments}


\begin{thebibliography}{99}
\bibitem{parity_eq} In Langevin dynamics, the detailed balance
 guarantees the parity symmetry, but it does not in Markovian jump dynamics~\cite{hklee,ckwon1,yeo}.
\end{thebibliography}
\end{document}